\documentclass[12pt,reqno]{amsart}
\usepackage{amsmath,amssymb,amsfonts,amsthm}
\usepackage[mathscr]{eucal}
\usepackage[all]{xy}
\usepackage{hyperref}
\usepackage{setspace}
\usepackage{bm}
\usepackage{xcolor}
\allowdisplaybreaks

\author{V.A.Abakumova, S.L.~Lyakhovich}

\address{Physics Faculty, Tomsk State University, Lenin ave. 36, Tomsk 634050, Russia.}

\email{abakumova@phys.tsu.ru, \, sll@phys.tsu.ru}

\title{Hamiltonian BFV-BRST Quantization for the Systems with Unfree Gauge Symmetry}

\begin{document}

\maketitle

\begin{abstract}
The BFV-BRST Hamiltonian quantization method is presented for the theories where the gauge  
parameters are restricted by differential equations. The general formalism is exemplified by the Maxwell-like theory of symmetric tensor field.
\end{abstract}
\section{Introduction}
\noindent The general formalism of gauge theories has been well-developed in details in recent decades, and it is considered nowadays as the textbook level knowledge.
Being a powerful uniform technology, the general gauge theory has the limitations of its own as it proceeds from certain assumptions that restrict the class of admissible models. In particular, the gauge parameters are assumed to be arbitrary functions. The unfree gauge symmetry, being the subject of our consideration, does not meet this assumption. The gauge parameters in this case are constrained by the differential equations.

The most known example of the unfree gauge symmetry is provided by the unimodular gravity (see, e.g. \cite{Percacci2018, BARVINSKY201759, PhysRevD.100.023542}, for discussion and various modifications), where the gauge parameters are restricted by the unimodularity condition, being the first-order differential equation. Among the other examples of the unfree gauge symmetry we can mention   Maxwell-like higher spin models \cite{Campoleoni2013}, and the higher spins with transverse symmetry \cite{SKVORTSOV2008301}.

Let us demonstrate the phenomenon of unfree gauge symmetry. Consider the theory whose equations of motion satisfy the identity 
\begin{equation}\label{id}
    \Gamma^i_\alpha\frac{\delta S}{\delta \varphi^i}\equiv -\Gamma_\alpha^a\tau_a, \quad \tau_a\neq K_a^i\frac{\delta S}{\delta \varphi^i},
\end{equation}
where $\tau_a$ are the completion functions, which do not reduce to a linear combination of Lagrangian equations, while they vanish on-shell, and $\Gamma_\alpha^a$ are differential operators with at maximum finite kernel,
\begin{equation}
    \Gamma_\alpha^au_a=0\,\, \Leftrightarrow \,\,u_a\in M=\text{Ker}\,\Gamma,\,\, \text{dim}\, M=n\in \mathbf{N}.
\end{equation}
The kernel $M$ is understood as a moduli space of the field theory, and its elements are parametrized by finite number of constant parameters $\Lambda$. The completion functions $\tau_a$ in (\ref{id}) are defined modulo the kernel $M$,
\begin{equation}
    \tau_a-u_a^A\Lambda_A\approx0,
\end{equation}
where specific values of the modular parameters are defined by asymptotics of the fields.

The action is invariant under the unfree gauge transformations
\begin{equation}
    \delta_\varepsilon\varphi^i=\Gamma_\alpha^i\varepsilon^\alpha,
\end{equation}
with the constraints on gauge parameters
\begin{equation}\label{constr}
    \Gamma_\alpha^a\varepsilon^\alpha=0.
\end{equation}
Indeed, 
\begin{equation}
    \delta_\varepsilon S \equiv \frac{\delta S}{\delta\varphi^i}\Gamma_\alpha^i\varepsilon^\alpha\equiv -\tau_a\Gamma_\alpha^a\varepsilon^\alpha=0,
\end{equation}
if the gauge parameters satisfy (\ref{constr}).

The Lagrangian structure of unfree gauge systems is established  in \cite{KAPARULIN2019114735}, for corresponding BV-BRST field-antifield formalism see \cite{Kaparulin2019}. In the present paper, we briefly demonstrate Hamiltonian structure of unfree gauge symmetry and BFV-BRST Hamiltonian quantization method proposed in \cite{Abakumova:2019uoo,Abakumova2020}. We exemplify the general formalism by the Maxwell-like theory of symmetric tensor field.

\section{Hamiltonian formalism}

\noindent Consider the Hamiltonian action of a theory with primary constraints,
\begin{equation}\label{SH}
    S_H=\int dt(p_i\dot{q}{}^i-H_T(q,p,\lambda)), \quad H_T(q,p,\lambda)=H(q,p)+\lambda^\alpha T_\alpha(q,p).
\end{equation}
The role of fields is played by canonical variables $q^i, p_i$, and Lagrange multipliers $\lambda^\alpha$. The summation over repeated indices includes integration over space.

According to the Dirac-Bergmann algorithm, the time derivative of the primary constraints reduces to the combination of primary and secondary constraints,
\begin{equation}\label{dotT}
    \dot{T}_\alpha=\{T_\alpha, H_T\}=V_\alpha^\beta T_\beta+\Gamma_\alpha^a\tau_a\approx0.
\end{equation}
Unfree gauge symmetry corresponds to the case when differential operator $\Gamma$ has at maximum finite kernel, or no inverse exists for $\Gamma$ in the class of differential operators.  
Completion functions $\tau_a$ can be redefined by adding modular parameters $\Lambda_a$ to make them vanishing on-shell,
\begin{equation}\label{dottau}
    \tau_a=\tau_a(q,p)-\Lambda_a\approx0, \quad \Lambda_a\in\text{Ker}\,\Gamma.
\end{equation}
The simplifying assumption is that no tertiary constraints appear,
\begin{equation}
    \dot{\tau}_a=\{\tau_a, H_T\}=V_a^\alpha T_\alpha+V_a^b\tau_b\approx0.
\end{equation}
For Hamiltonian formalism for the systems with tertiary and higher order constraints see \cite{PhysRevD.102.125003}. 
Unfree gauge transformations for constrained Hamiltonian system read
\begin{eqnarray}\label{transfH}
    \delta_\varepsilon O(q,p)=\{O,T_\alpha\}\varepsilon^\alpha&&+\{O,T_a\}\varepsilon^a,\quad
    \delta_\varepsilon\lambda^\alpha=\dot{\varepsilon}^\alpha+V_\beta^\alpha\varepsilon^\beta+V_a^
    \alpha\varepsilon^a,\\\label{constrH}
    &&\dot{\varepsilon}^a+V_b^a\varepsilon^b+\Gamma_\alpha^a\varepsilon^\alpha=0.
\end{eqnarray}
It can be verified by direct variation, that the action (\ref{SH}) is invariant under the transformations (\ref{transfH}) if gauge parameters are subject to the constraints (\ref{constrH}),
\begin{equation}
    \delta_\varepsilon S_H\equiv\int dt\big((\dot{\varepsilon}^a+V_b^a\varepsilon^b+\Gamma_\alpha^a\varepsilon^\alpha)\tau_a-\frac{1}{2}\frac{d}{dt}(T_\alpha\varepsilon^\alpha+\tau_a\varepsilon^a)\big).
\end{equation}
The gauge identities read
\begin{eqnarray}\label{GIH1}
    &&\{T_\alpha,q^i\}\frac{\delta S}{\delta q^i}+\{T_\alpha, p_i\}\frac{\delta S}{\delta p_i}+\big(\delta_\alpha^\beta\frac{d}{dt}-V_\alpha^\beta\big)\frac{\delta S}{\delta \lambda^\beta}+\Gamma_\alpha^a\tau_a\equiv0;
    \\\label{GIH2}
    &&\{\tau_a,q^i\}\frac{\delta S}{\delta q^i}+\{\tau_a,p_i\}\frac{\delta S}{\delta p_i}-V_a^\alpha\frac{\delta S}{\delta \lambda^\alpha}+\big(-\delta_a^b\frac{d}{dt}+\Gamma_a^b\big)\tau_b\equiv0.
\end{eqnarray}

Notice natural ambiguity in the definition of the constraints and Hamiltonian. The Hamiltonian can be modified by adding constraint terms, while the constraints of all generations can be linearly combined. These modifications result in equivalent gauge identities (\ref{GIH1}), (\ref{GIH2}). Hence, these modifications do not change gauge symmetry.

\section{Hamiltonian BFV-BRST formalism}

\noindent The minimal sector is introduced along the same lines as for any first-class constrained system. Every first-class constraint is assigned with a canonical pair of ghosts,
\begin{equation}
    \{C^\alpha, \bar{P}_\beta\}=\delta^\alpha_\beta, \quad \text{gh}(C^\alpha)=-\text{gh}(\bar{P}_\alpha)=1; \quad
    \{C^a, P_b\}=\delta^a_b, \quad \text{gh}(C^a)=-\text{gh}(\bar{P}_a)=1.
\end{equation}
Hamiltonian BFV-BRST generator of minimal sector begins with constraints,
\begin{equation}
    Q_{\text{min}}=C^\alpha T_\alpha+C^a\tau_a+\ldots, \quad \text{gh}(Q_{\text{min}})=1, \quad \{Q_{\text{min}},Q_{\text{min}}\}=0,
\end{equation}
where $\ldots$ stands for $\bar{P}$-depending terms. The ghost extension of the Hamiltonian begins with original Hamiltonian $H$,
\begin{equation}
    \mathcal{H}=H(q,p)+\ldots, \quad \text{gh}(\mathcal{H})=0, \quad \{Q_{\text{min}},\mathcal{H}\}=0.
\end{equation}

The non-minimal sector is asymmetric with the minimal one unlike the usual BVF-formalism. The number of gauge-fixing conditions coincides with the number of primary constraints, so the same number of non-minimal sector ghosts is introduced,
\begin{equation}
    \{P^\alpha,\bar{C}_\beta\}=\delta^\alpha_\beta, \quad \text{gh}(P^\alpha)=-\text{gh}(\bar{C}_\alpha)=1.
\end{equation}
The non-minimal sector also includes the Lagrange multipliers $\lambda^\alpha$ to the primary constraints, and the Lagrange multipliers $\pi_\alpha$ to the independent gauge conditions $\dot{\lambda}^\alpha-\chi^\alpha(q,p)=0$,
\begin{equation}
    \{\lambda^\alpha, \pi_\beta\}=\delta^\alpha_\beta, \quad \text{gh}(\lambda^\alpha)=\text{gh}(\pi_\alpha)=0.
\end{equation}
The complete BRST-charge reads
\begin{equation}
    Q=Q_{\text{min}}+\pi_\alpha P^\alpha,
\end{equation}
and the gauge-fixed Hamiltonian is defined by the usual rule,
\begin{equation}
    H_\Psi=\mathcal{H}+\{Q,\Psi\}, \quad \Psi=\bar{C}_\alpha\chi^\alpha+\lambda^\alpha\bar{P}_\alpha.
\end{equation}
Given the Hamiltonian $H_\Psi$, it defines gauge-invariant quantum dynamics.

\section{Example: Maxwell-like theory of symmetric tensor field}

\noindent Consider the symmetric traceful second-rank tensor field $h_{\mu\nu}(x), h^\mu{}_\mu\equiv h$ in $d$-dimensional Minkowski space with mostly negative metric $\eta_{\mu\nu}=\text{diag}(1,-1,\ldots,-1)$. The Maxwell-like Lagrangian \cite{Campoleoni2013} reads
\begin{equation}\label{LMl}
    \mathcal{L}=\frac{1}{2}\big(\partial_\mu h_{\nu\rho}\partial^\mu h^{\nu\rho}-2\partial^\mu h_{\mu\rho}\partial_\nu h^{\nu\rho}\big)+\partial_\mu(h_{\mu\rho}\partial_\nu h^{\nu\rho})+A\Lambda h+B\Lambda^2h^2, \quad A,B,\Lambda=const.
\end{equation}
For the sake of generality, we consider some additional terms compared to \cite{Campoleoni2013}, which do not change the gauge symmetry. The Lagrangian equations for (\ref{LMl}) read
\begin{equation}
    \frac{\delta S}{\delta h^{\mu\nu}}\equiv-\square h_{\mu\nu}+\partial_\mu\partial^\rho h_{\rho\nu}+\partial_\nu\partial^\rho h_{\rho\mu}=0,
\end{equation}
and gauge identities (\ref{id}) have the form
\begin{equation}
    2\partial^\nu \frac{\delta S}{\delta h^{\mu\nu}}=\partial_\mu \tau, \quad \tau=2\partial^\mu\partial^\nu h_{\mu\nu}.
\end{equation}
This relation means that $\tau$ reduces on-shell to the element of the kernel of $\partial_\mu$,  so it is just a constant on-shell,
\begin{equation}
    \partial_\mu\tau=0 \,\, \Rightarrow \,\, \tau-C\Lambda\approx 0, \quad C, \Lambda=const.
\end{equation}
The unfree gauge transformations read
\begin{equation}\label{constrMl}
    \delta_\varepsilon h_{\mu\nu}=\partial_\mu\varepsilon_\nu+\partial_\nu\varepsilon_\mu, \quad  \partial_\mu\varepsilon^\mu=0.
\end{equation}

The corresponding Hamiltonian action reads
\begin{eqnarray}\label{SHMl}
    S_H&=&\int d^dx \big(\Pi_{ij}\dot{h}^{ij}+\Pi\dot{\lambda}-H_T\big), \quad H_T=H+\lambda^iT_i, \quad T_i=-2(\partial^j\Pi_{ji}-\partial_i\Pi),\\\label{HMl}
    H&=&\frac{1}{2}(\Pi_{ij}\Pi^{ij}-\Pi^2)-\frac{1}{2}(\partial_ih_{jk}\partial^ih^{jk}+\partial_i\lambda\partial^i\lambda)+\partial^jh_{ji}\partial_kh^{ki}\\\notag
    &-&A\Lambda(\lambda+h^i{}_i)-B\Lambda^2(\lambda^2+2\lambda h^i{}_i+h^i{}_ih^j{}_j),
\end{eqnarray}
where $\lambda\equiv h^{00}, \lambda^i\equiv h^{0i}$, $h^i{}_i=\eta^{ij}h_{ij}$. Conservation of primary constraints $T_i$ leads to the secondary one,
\begin{equation}
    \dot{T}_i=\{T_i,H_T\}=-\partial_i\tau_0, \quad
    \tau_0=2(\partial^i\partial^jh_{ij}+\Delta\lambda)-C\Lambda,
\end{equation}
and no tertiary constraints appear,
\begin{equation}
    \dot{\tau}_0=\{\tau_0,H_T\}=-\partial^iT_i.
\end{equation}
Unfree gauge symmetry transformations read
\begin{eqnarray}
    &\delta_\varepsilon h_{ij}=\partial_i\varepsilon_j+\partial_j\varepsilon_i, \quad &\delta_\varepsilon\lambda=-2\partial_i\varepsilon^i,\\
    &\delta_\varepsilon\lambda^i=\dot{\varepsilon}^i+\partial^i\varepsilon^0, \quad &\dot{\varepsilon}^0+\partial_i\varepsilon^i=0.
\end{eqnarray}

Consider Hamiltonian BFV-BRST formalism for the Maxwell-like theory. Complete BRST charge reads
\begin{equation}
    Q=-2C^i\big(\partial^j\Pi_{ji}-\partial^i\Pi\big)+C^0\big(2(\partial^i\partial^jh_{ij}+\Delta\lambda)-C\Lambda\big)+\pi_iP^i.
\end{equation}
Given independent gauge-fixing conditions
\begin{equation}
    \dot{\lambda}^i-\chi^i=0, \quad \chi^i=-\partial_jh^{ji},
\end{equation}
the gauge-fixed Hamiltonian is introduced,
\begin{equation}
    H_\Psi=H-C^0\partial^i\bar{P}_i-C^i\partial_i\bar{P}_0+\lambda^iT_i+\partial_j\bar{C}_i\partial^jC^i+\partial^i\bar{C}_i\partial_jC^j-\pi_i\partial_jh^{ji}+\bar{P}_iP^i,
\end{equation}
where $H$ is defined by (\ref{HMl}). The corresponding path integral reads
\begin{eqnarray}\label{ZpsiMl}
    Z_\Psi&&=\int[D\Phi]\exp\big\{\frac{i}{\hbar}\int d^dx\big[\Pi^{ij}\dot{h}_{ij}+\Pi\dot{\lambda}+\pi_i(\dot{\lambda}^i+\partial_jh^{ji})-H-\lambda^iT_i\\\nonumber
    &&+\,\bar{C}_i(\Delta C^i+\partial^i\partial_jC^j)+\bar{P}_0(\dot{C}^0+\partial_iC^i)-(\bar{P}_i+\dot{\bar{C}}_i)P^i+\bar{P}_i(\dot{C}^i+\partial^iC^0)\big]\big\},
\end{eqnarray}
where $\Phi=\{h_{ij},\Pi^{ij},\lambda, \Pi, \lambda^i,\pi_i,C^i,\bar{P}_i, C^0,\bar{P}_0,P^i,\bar{C}_i\}$. Integrating (\ref{ZpsiMl}) over $\Pi^{ij}, \Pi, \bar{P}_i,P^i$, we get its Lagrangian representation
\begin{equation}
    Z=\int[D\Phi']\exp\big\{\frac{i}{\hbar}\int d^dx\big[\mathcal{L}+\pi_i(\partial_\mu h^{\mu i})+\bar{C}_i(\square\delta_\mu^i+\partial^i\partial_\mu)C^\mu+\bar{P}_0\partial_\mu C^\mu \big]\big\},
\end{equation}
where $\Phi'=\{h_{\mu\nu},\pi_i,C^\mu,\bar{P}_0,\bar{C}_i\}$, and $\mathcal{L}$ is the original Lagrangian (\ref{LMl}). Note that constraint on ghosts mirrors transversality constraint (\ref{constrMl}) imposed on gauge parameters.

Let us consider Maxwell-like Lagrangian (\ref{LMl}) with specific cubic vertex \cite{Francia2017},
\begin{equation}\label{Lint}
    \mathcal{L}^{int}=\mathcal{L}+h_{\mu\nu}h^{\mu\nu}\tau,
\end{equation}
where $\tau=2\partial^\mu\partial^\nu h_{\mu\nu}-C\Lambda$ is a completion function of free theory. 
By Legendre transform of (\ref{Lint}), and using the change of variables
\begin{equation}\label{varH}
    \lambda \rightarrow \lambda-2g(\lambda^2+2\lambda_i\lambda^i+h_{ij}h^{ij}), \quad C \rightarrow C-A,
\end{equation}
we get the Hamiltonian action
\begin{equation}\label{SHint}
    S_H^{int}=\int d^dx \big(\Pi_{ij}\dot{h}^{ij}+\Pi\dot{\lambda}-H_T^{int}\big), \quad H_T^{int}=H_T-g(\lambda^2+2\lambda_i\lambda^i+h_{ij}h^{ij})\tau_0,
\end{equation}
where $H_T, \tau_0$ are total Hamiltonian and secondary constraint of free Maxwell-like theory. 
The primary and secondary constraints for the interacting theory read
\begin{equation}
    T_i^{int}=T_i+4g\lambda_i\tau_0, \quad \tau_0^{int}=\tau_0,
\end{equation}
where $T_i$ and $\tau_0$ are primary and secondary constraints for the free theory.
So, the inclusion of vertex (\ref{Lint}) leads to the linear combination of Hamiltonian constraints, modification of the Hamiltonian by adding constraint terms, and the local change of variables (\ref{varH}). This means that the vertex (\ref{Lint}) results in the equivalence transformation of the free theory (\ref{LMl}).

\section{Conclusion}
\noindent We explain how the BFV-BRST quantization method is adjusted for the case of unfree gauge symmetry. The main modifications are related to the non-minimal sector of ghosts. As an example, we consider the Maxwell-like theory whose Hamiltonian formalism  has not been addressed elsewhere. We demonstrate the triviality of cubic vertex for this theory from the viewpoint of Hamiltonian formalism.

\vspace{0.4 cm} \noindent \textbf{Acknowledgments.} The study of the Hamiltonian structure of unfree gauge symmetry was supported by a government task of the Ministry of Science and Higher Education of the Russian Federation, Project No. 0721-2020-0033. The application of the general formalism to the Maxwell-like theory of symmetric tensor field was supported by the Foundation for the Advancement of Theoretical Physics and Mathematics "BASIS".

\end{document}